# Status Update and Closed-Loop Performance of the Magellan Adaptive Optics VisAO Camera

Derek Kopon[*][a], Laird M. Close[a], Jared Males[a], Victor Gasho[a], Katie Morzinski[a], Katherine Follette[a]
[a]CAAO, Steward Observatory, University of Arizona, Tucson AZ USA 85721

## ABSTRACT

We present laboratory results of the closed-loop performance of the Magellan Adaptive Optics (AO) Adaptive Secondary Mirror (ASM), pyramid wavefront sensor (PWFS), and VisAO visible adaptive optics camera. The Magellan AO system is a 585-actuator low-emissivity high-throughput system scheduled for first light on the 6.5 meter Magellan Clay telescope in November 2012. Using a dichroic beamsplitter near the telescope focal plane, the AO system will be able to simultaneously perform visible (500-1000 nm) AO science with our VisAO camera and either 10 μm or 3-5 μm science using either the BLINC/MIRAC4 or CLIO cameras, respectively. The ASM, PWS, and VisAO camera have undergone final system tests in the solar test tower at the Arcetri Institute in Florence, Italy, reaching Strehls of 37% in i'-band with 400 modes and simulated turbulence of 14 cm $r_o$ at v-band. We present images and test results of the assembled VisAO system, which includes our prototype advanced Atmospheric Dispersion Corrector (ADC), prototype calcite Wollaston prisms for SDI imaging, and a suite of beamsplitters, filters, and other optics. Our advanced ADC performs in the lab as designed and is a 58% improvement over conventional ADC designs. We also present images and results of our unique Calibration Return Optic (CRO) test system and the ASM, which has successfully run in closed-loop at 1kHz. The CRO test is a retro reflecting optical test that allows us to test the ASM off-sky in close-loop using an artificial star formed by a fiber source.

## 1. INTRODUCTION: MAGELLAN SIMULTANEOUS VISIBLE AND IR ADAPTIVE OPTICS

The Magellan Clay telescope (Fig. 1) is a 6.5m Gregorian telescope located at the Las Campanas Observatory (LCO) in Chile. In the fall of 2012 we, along with our collaborators at the Arcetri Observatory and the Carnegie Institute, will be commissioning the first Adaptive Optics (AO) system on the Magellan telescope. Our AO system uses an 85 cm diameter adaptive secondary mirror (ASM) with 585 actuators. With the exception of our smaller number of actuators and ASM diameter (585 actuators vs. 672 on LBT), our system is a virtual clone of the recently commissioned LBT AO system (Esposito et al. 2010), using the same secondary optical prescription, control hardware, software, and pyramid wavefront sensor (PWFS). Because our primary mirror is slightly smaller than the LBT 8m mirror, we have finer on-sky actuator spacing and therefore expect excellent Strehls in the IR (~90% at H-band), and decent Strehls in the visible (~40% at 770 nm).

To take advantage of high-Strehl AO correction and the excellent seeing at the Magellan site ($r_o$ frequently as high as 20 cm at 550 nm) our first light commissioning will make use of two science instruments: the CLIO 3-5 μm camera (see Sivanandam et al. 2006) and our VisAO camera (Kopon et al. 2009). While the original motivation for an adaptive secondary system is to reduce thermal background by reducing the number of warm reflections in the optical train, the fine on-sky actuator pitch and fast actuator response time will also allow good correction at shorter wavelengths. Our dichroic beamsplitter located near the telescope focal plane will transmit IR light to CLIO and will reflect visible light to the W-unit, an optical board containing both the PWFS and VisAO camera. With our predicted ~40% Strehl and 8.5 mas pixels, we plan to make images > 4.7 times better than the highest resolution instrument on HST (none of the instruments currently operational on HST Nyquist sample at less than 1 μm) over our 8.5 arcsec FOV. A suite of filters, beamsplitters, and other optical components will allow many different observing modes in the visible depending on the science case and the guide star magnitude.

At the time of this writing, all of the hardware has arrived in Chile after passing our preship review and final system testing in Florence, Italy. The full system, with all of the VisAO optical components, the ASM, and the PWFS was tested closed loop at 1 kHz in the test tower. The hardware has been unpacked and will be installed on the telescope in November of this year.

---

[*] dkopon@as.arizona.edu

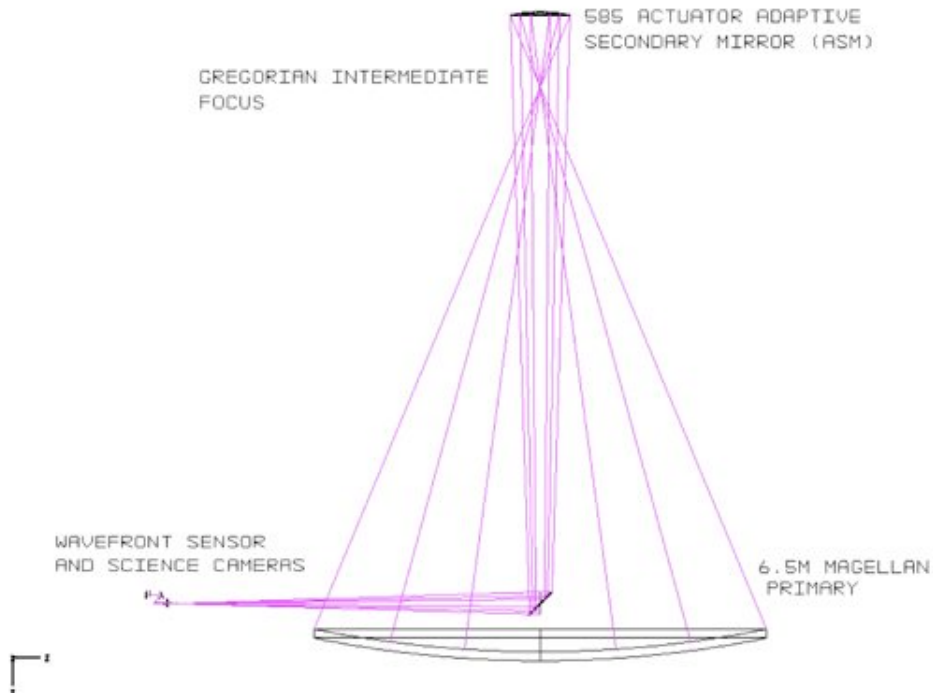

**Figure 1:** Raytrace of the Magellan telescope with the F/16 adaptive secondary. The ASM is a concave ellipsoidal mirror with an intermediate focus at the near ellipsoidal conjugate where a retro-reflecting optic can be placed to enable off-sky testing.

## 2. THE W-UNIT: PYRAMID WAVEFRONT SENSING AND THE VISAO CAMERA

Visible light reflected from the dichroic travels to the W-unit, an optical board mounted on a precision 3-axis translation stage that can patrol a 2.3x3.2 arcmin field to acquire guide stars and science targets. The W-unit contains both the PWFS and the VisAO science camera. The ability to move around this 2.3x3.2 arcmin field, combined with a telecentric lens that corrects field curvature, allows us to look several minutes away from the guide star with the CLIO IR camera, thereby significantly increasing our sky coverage. The visible light reflected from the dichroic to the W-unit passes through a triplet lens that changes the beam from a diverging F/16 beam to a converging F/49 beam. The light then passes through the triplet ADC before reaching the beamsplitter wheel that contains a host of beamsplitters and dichroics. The transmitted light from the B/S wheel goes to the PWFS and the reflected light goes to the VisAO science camera. Non-common path errors are minimized by having the beamsplitter wheel as late as possible in the optical train. Figure 2 shows the layout of the W-unit.

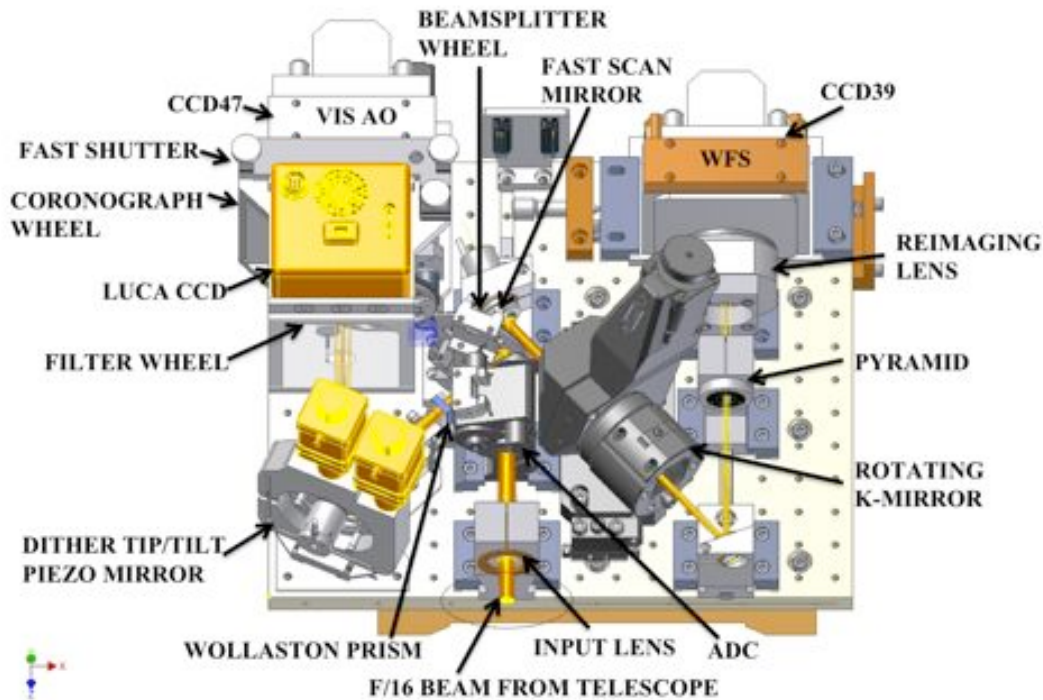

**Figure 2:** The W-unit optical board. The board is mounted on a precision 3-axis translation stage that can patrol a 2.3'x3.2' field in order to acquire guide stars and science targets. Incoming light from the F/16 telescope focus passes though an input lens and the ADC before hitting the beamsplitter/dichroic wheel. Light transmitted by the beamsplitter goes to the PWFS channel. Light reflected from the beamsplitter goes to the VisAO science channel.

### 2.1 Pyramid Wavefront Sensor

Transmitted light from the beamsplitter wheel enters the PWFS channel and reflects off of a piezo fast steering mirror located near the reimaged pupil that is used to circularly modulate the guide star image around the pyramid tip. Next, the light passes through a K-mirror rerotator that rotates the pupil so that the image of the ASM actuators stays fixed with respect to the CCD39. A double pyramid followed by a reimaging lens then form four pupil images in the quadrants of the CCD39. The resultant image on the CCD39 is four pupil images whose intensity variations can be used to reconstruct the wavefront. A detailed description of the operation of the pyramid sensor (PS) arm of the W-unit can be found in Esposito et al. 2008. The pyramid sensor is very important for visible AO because of its potential for approaching a telescope's diffraction limit and the dynamic range that is a result of being able to change the binning of the CCD39. A Shack-Hartmann (SH) sensor is diffraction limited by the size of a pupil sub-aperture: i.e. the pitch of the lenslet array. Whereas, the PS uses the full pupil aperture and is only diffraction limited by the size of the primary mirror. Since the wavefront sensing wavelengths are essentially the same as the science wavelengths (~0.7 $\mu$m), it is essential that the wavefront sensor be as close to the diffraction limit as possible (Esposito et al. 2000). The dynamic range provided by the PS should allow us to use ~2 mag fainter guide stars than would be allowed by the SH sensor, in addition to allowing us to come very close to diffraction limited correction (Esposito et al. 2001).

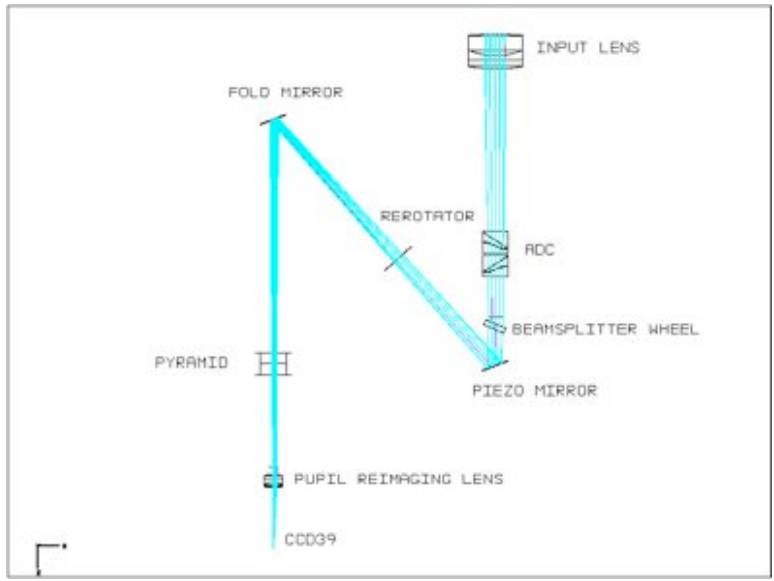

**Figure 3:** The optical path of the PWFS channel of the W-unit. The guide star is imaged onto and modulated around the tip of the pyramid, thereby forming 4 pupil images in the quadrants of the CCD39.

## 2.2 Advanced Triplet ADC

While atmospheric dispersion is generally not a significant effect in the IR over the wavebands at which we are observing, the effect is quite significant at visible wavelengths. There is great scientific incentive to push into the shorter wavelengths and to be able to observe over a wide band in order to see faint companions and circumstellar structure. Over the band 500-1000 nm, atmospheric dispersion creates a PSF that is 2000 μm long in the dispersed direction and diffraction limited (30-60 μm) in the other. Correcting this large amount of dispersion over a large band requires an ADC that can correct both primary and secondary chromatism (see Kopon et al. 2009). In addition to correcting the atmospheric dispersion for the VisAO science imaging, the correction is also important for the PWFS. Having a tight, well-corrected image of the guide star on the pyramid tip allows for tighter modulation radii and better AO correction.

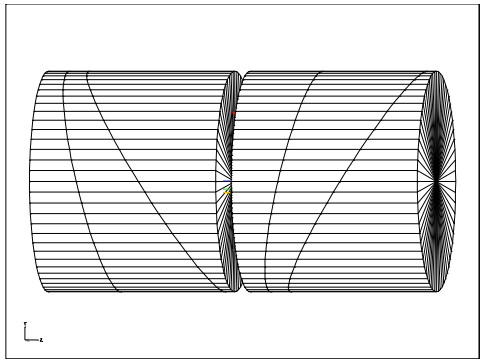

**Figure 4:** The advanced triplet ADC. The two counter-rotating triplets correct both primary and secondary color out to 70° zenith angle.

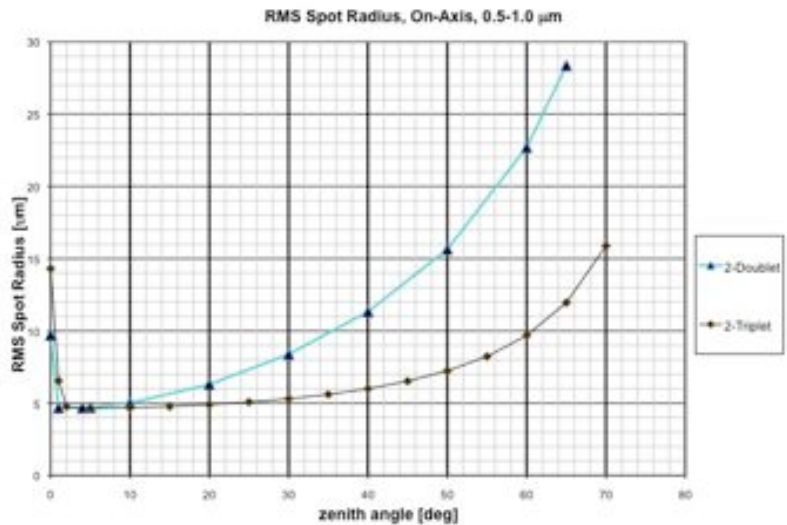

**Figure 5:** RMS spot size as a function of zenith angle showing the performance of a conventional ADC design and our novel 2-triplet design. Our design performs 58% better at higher zenith angles.

We designed, fabricated, and tested a triplet ADC that corrected both primary and secondary chromatism and allows us to reach the broadband diffraction limit out to a Zenith angle of 70 degrees (2.9 air masses). We tested the dispersion and image quality of these optics using a special lab setup consisting of a white light fiber source and several narrow band filters (see Kopon et al. 2010). Our ADC was adopted by our Arcetri collaborators for use on the LBT AO system. Its performance was recently verified on sky at 30 degree zenith angle with the AO system running in closed loop (see Kopon et al. 2012).

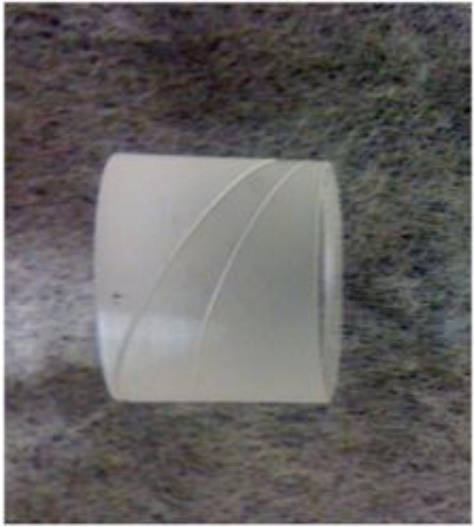

**Figure 6:** An as fabricated ADC prism.

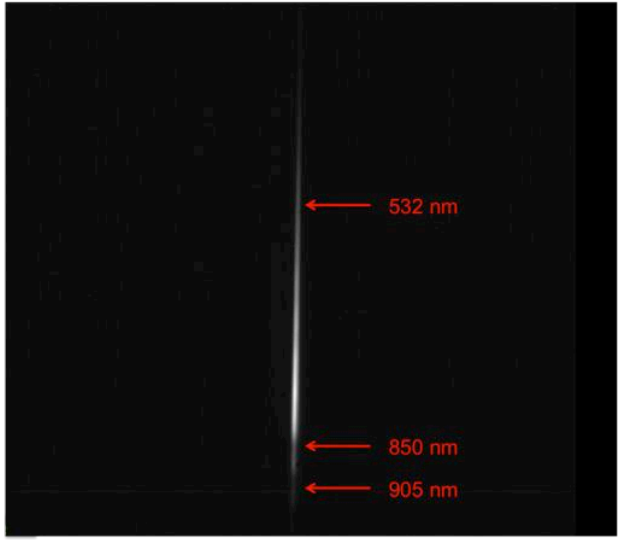

**Figure 7:** The dispersed white light source used to test the ADC in the lab.

### 2.3 VisAO Optics

After passing through the ADC, the light reaches a beamsplitter wheel whose components are listed in Table 1. Light reflected from the beamsplitter or dichroic then continues in the VisAO channel where it has the option of passing through a removable Wollaston prism for Spectral Differential Imaging (SDI) (Close et al. 2005). The light then reflects off of a silver tip/tilt dither mirror that can be used to move the science target around the CCD47, for example to align a bright central star to the center of a coronographic spot. Following the dither mirror, there is a filter wheel containing a suit of circular Sloan filters and a long-pass filter (Table 1). Following the sloan wheel is a baffle tube and then another wheel, which we call the coronograph wheel. This wheel holds the SDI filter sets and various coronographic spot patterns that can be switched in and out passed on the science goals.

The last component before the CCD47 is our fast asynchronous shutter. This shutter can serve several purposes. The first is to limit the exposure time for bright objects. For example, our spatial resolution is such that we expect to be able to resolve features on the surfaces of nearby super giants, such as Betelgeuse. A science target this bright will require the fast shutter in order to avoid saturation. A second use of the fast shutter will be for real-time frame selection (RTFS), a technique for blocking low-Strehl light from reaching the CCD47. Simulations of our AO performance using CAOS, an AO simulation code, show that our Strehls in the visible will vary significantly over several hundred ms from Strehsl as high as 50% to as low as 10%. By using real-time telemetry to control the fast shutter, we can block the light during periods of low Strehl from reaching the CCD47. We can therefore achieve the improved resolution of frame selection, combined with the low-read noise of a long integration, since the CCD47 is not being read out during this time. For more details on the fast shutter and RTFS, see Males et al. these proceedings.

| Position | Beamsplitter | Filter Wheel | Coronagraph |
| --- | --- | --- | --- |
| 1 | Bare glass (96/4) | SDSS z' | Hα SDI |
| 2 | 50/50 | SDSS i' | [OI] SDI |
| 3 | Dichroic (reflect λ < 800 nm) | SDSS r' | [SII] SDI |
| 4 | Dichroic (reflect λ < 950 nm) | open | Coronographic spot pattern "#5" |
| 5 | Aluminum puck for taking "darks" | Long-pass "y" (λ > 950 nm) | Coronographic spot pattern "#6" |
| 6 | ----------------- | -------------- | open |

**Table 1:** VisAO optical components

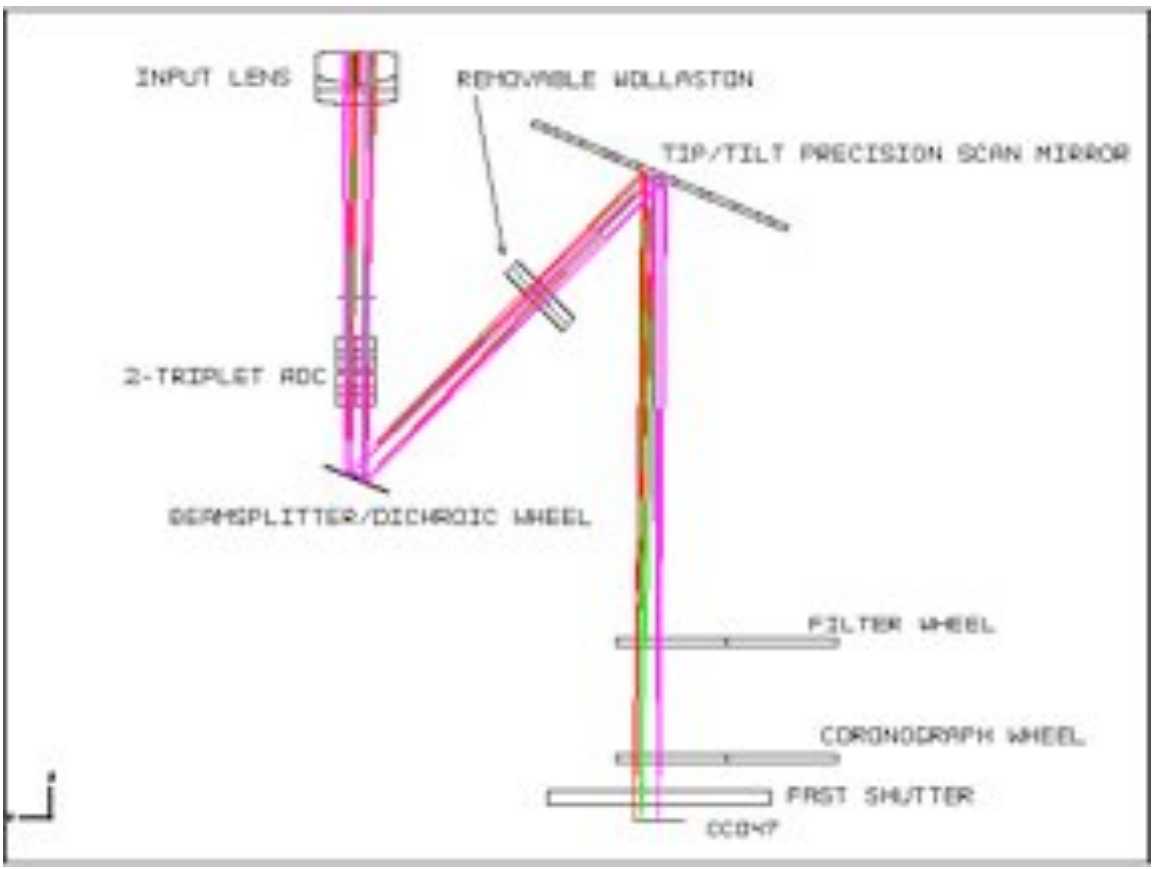

**Figure 8:** The VisAO optical path.

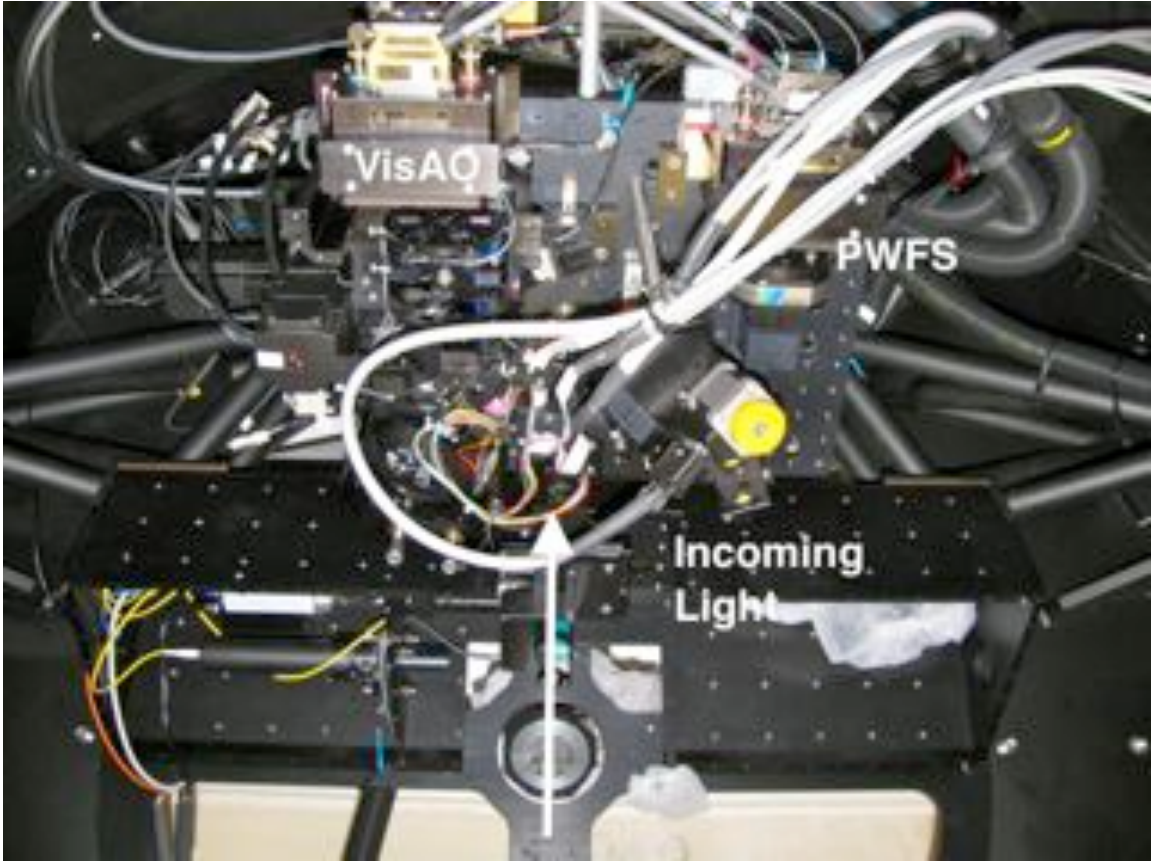

**Figure 9:** The fully assembled W-unit installed in the NAS in Arcetri just before tower testing.

**2.4 SDI High-Resolution Imaging**

SDI is a powerful form of speckle suppression that operates by splitting the incoming beam using a low-angle (~1.5 degree) Wollaston prism. This method of beam splitting is useful for high-resolution imaging because it allows both beams to continue on the same optical path, thereby minimizing non-common-path errors to ~10 nm. By the time the two images reach the focal plane, they have been deviated by ~6 mm, roughly half the size of the focal plane, so they can be imaged in different halves of the same detector. Just before the CCD47, two narrow-band filters can be placed, each covering their respective halves of the focal plane. One of these filters is centered on some wavelength of scientific interest (H$\alpha$ for example) and the other is center just off of this wavelength at some continuum value. These two images are then subtracted from each other and broadband specs are mitigated, revealing whatever faint companions or structures are strongly emitting the spectral line of interest. For more on specific SDI science cases, see Follette et al. 2010 or Close et al. these proceedings.

Our Wollaston prism is located on an elevator platform that can raise and lower the prism in and out of the beam. Our initial prototype Wollaston was a custom quartz prism made by Halle in Germany. Because of the low birefringence of quartz, the prism required a steep (~40 degree) cut angle in order to achieve our required ~1.5 degree beam deviation angle. This steep cut angle introduced too much astigmatism into the beam. We then proceeded with a prototype calcite prism made by Leysop. Because of the much higher birefringence of calcite, the shallow cut angle resulted in much lower astigmatism and better image quality.

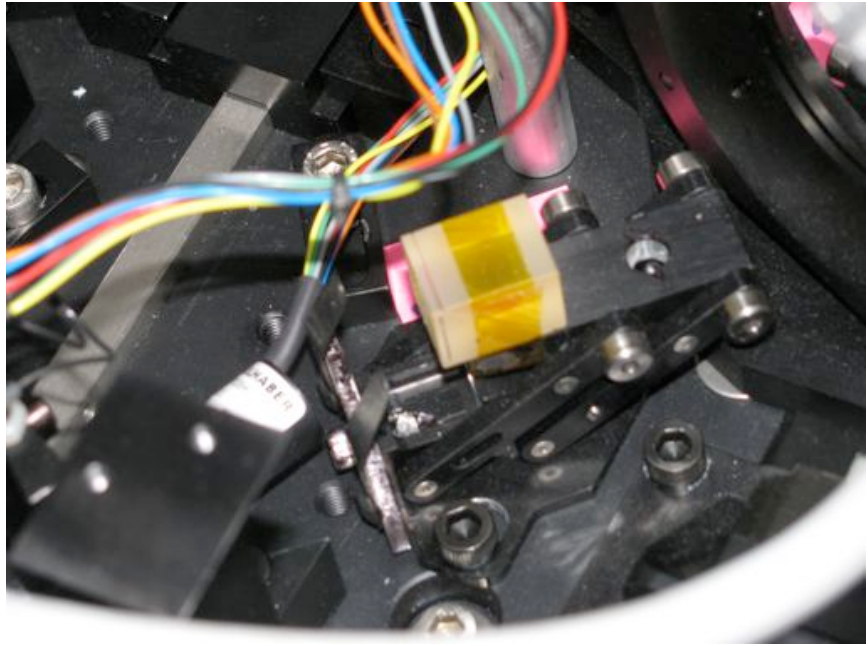

**Figure 11:** The prototype quartz Wollaston mounted in the W-unit on its elevator platform for raising and lowering into and out of the beam.

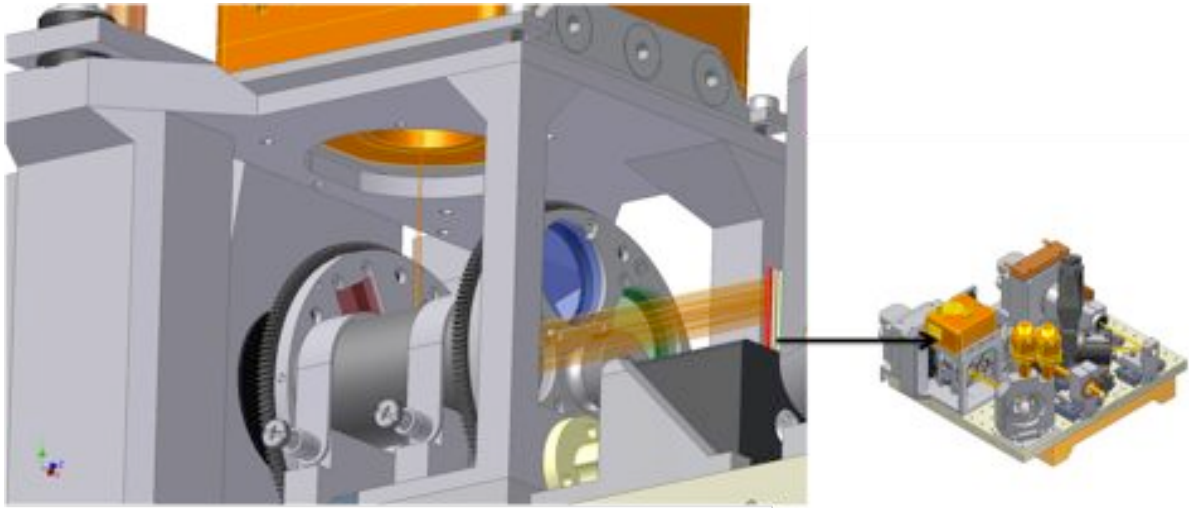

**Figure 12:** Close up of the SDSS filter wheel, the coronograph/SDI filter wheel, and the CCD47.

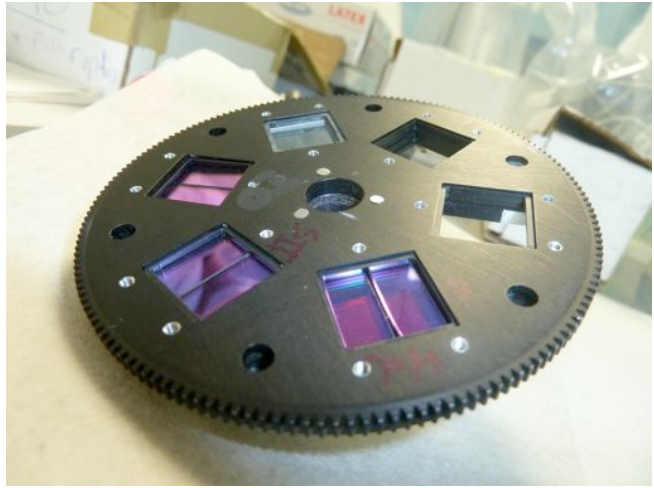

**Figure 13:** The coronograph wheel with the SDI filter pairs and one of the chrome coronographic spots.

**2.5 Coronographic Spots**

The high spatial resolultion and limited FOV of the VisAO camera necessarily mean that most of our science cases will be circumstellar structures (disks, jets, etc.) and companions. Because of this, there is a need to block the bright light of the central star in order to avoid CCD saturation and blooming. To that end, we have designed a series of chrome-on-glass masks that have been optimized for various science cases. For more on chrome-on-glass coronographic spots, see Park et al. 2006.

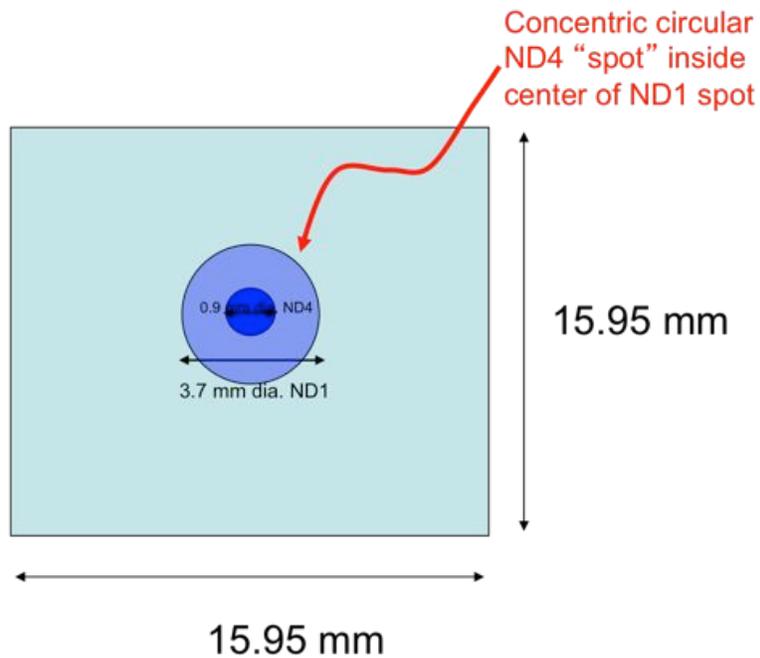

**Figure 14:** One of our custom chrome-on-glass coronographic spot designs. r = 0.1 arcsec ND4 and 0.1 < r < 1.0 arcsec ND1.

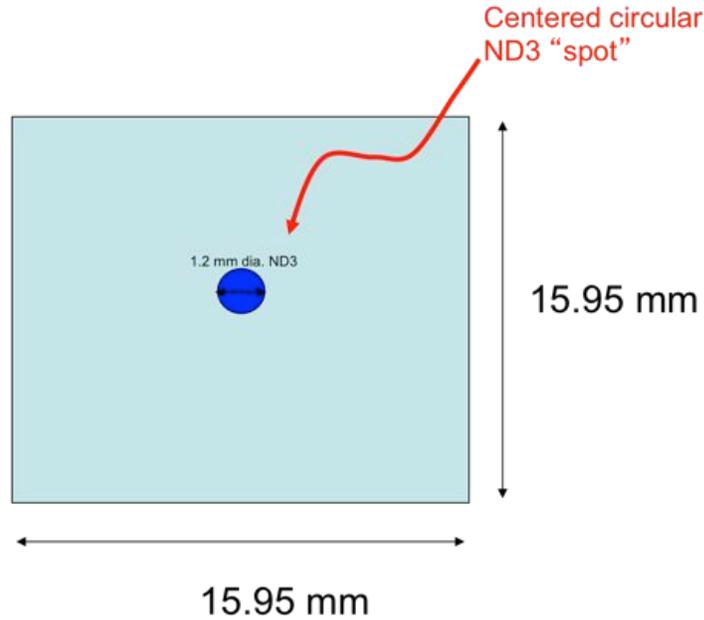

**Figure 14:** Another chrome spot pattern. r = 0.2 arcsec ND3 chrome on glass.

## 3. THE ARCETRI TOWER TESTS

From the summer of 2011 through the spring of 2012, the Magellan AO system (the ASM, W-unit, and other hardware) was in Florence, Italy undergoing testing in the Arcetri solar test tower. The Arcetri test tower is tall enough (14 m) to allow the ASM to be tested at the same focal distance that it will be used at the Magellan telescope. The ASM is mounted at the top of the tower and either a 4D interferometer or the W-unit is mounted at the base. Because the ASM is a concave ellipsoid, as required by Magellan's Gregorian design, a retro-reflecting optic, which we call the Calibration Return Optic (CRO), is mounted at the short ellipsoidal conjugate. This enables a double-pass optical test to be performed.

The 4D interferometer is a simultaneous phase-shifting interferometer that allows interferometric measurements of the secondary surface to be made quickly and reliably in a noisy environment. This setup is used initially to "flatten" the ASM, i.e. to calibrate the actuator forces required to shape the ASM shell into its nominal ellipsoidal shape and to measure the actuator influence functions. The 4D is then replaced with the NAS (our name for our large mechanical support structure) that holds the W-unit and a white-light fiber source calibration unit called the "flower pot." The flower pot fiber is used to stimulate a star that is reflected off of the ASM, then the CRO, then back to the ASM, and finally reimaged at the F/16 focal plane before entering the W-unit, which contains the PWFS and the fully integrated VisAO camera. This setup allows us to run the AO system in closed loop. Turbulence can be simulated using the ASM itself by injecting a pre-calculated phase screen in addition to the AO corrections. This phase screen is designed to simulate a conservative seeing of the Magellan site (FWHM of 0.8" at $\lambda$ = 550 nm, or $r_o$ = .14 m). For more on how turbulence is simulated with the ASM, see Esposito et al. 2010.

### 3.1 The Calibration Return Optic Setup

In order to test the ASM and AO system closed loop off-sky, a retro-reflecting optic called the Calibration Return Optic (CRO) is place at the short ellipsoidal conjugate of the ASM. This CRO operates in double pass and consists of an on-axis parabolic mirror that reflects incoming light to a return flat and then projects the beam back to the secondary. The simulated star is a broad band white light fiber source located in the "flower pot" near the F/16 focal plane. This light reflected off of a beamsplitter cube located at the F/16 focal and sent to the ASM. The return beam is then transmitted through the beamsplitter cube to the W-unit, which contains the PWFS and VisAO camera that can be used a science camera or technical viewer. We have designed this test setup so that we can use it at the Magellan telescope in order to test or periodically recalibrate the system off sky.

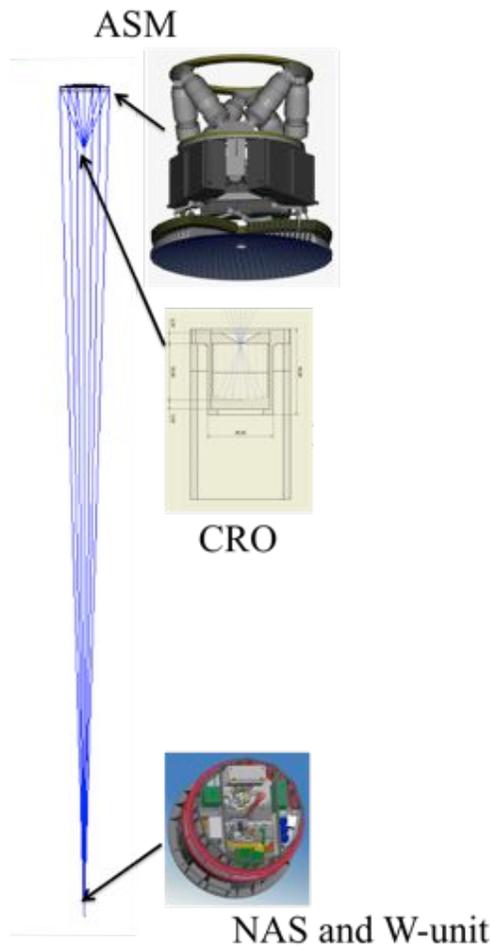 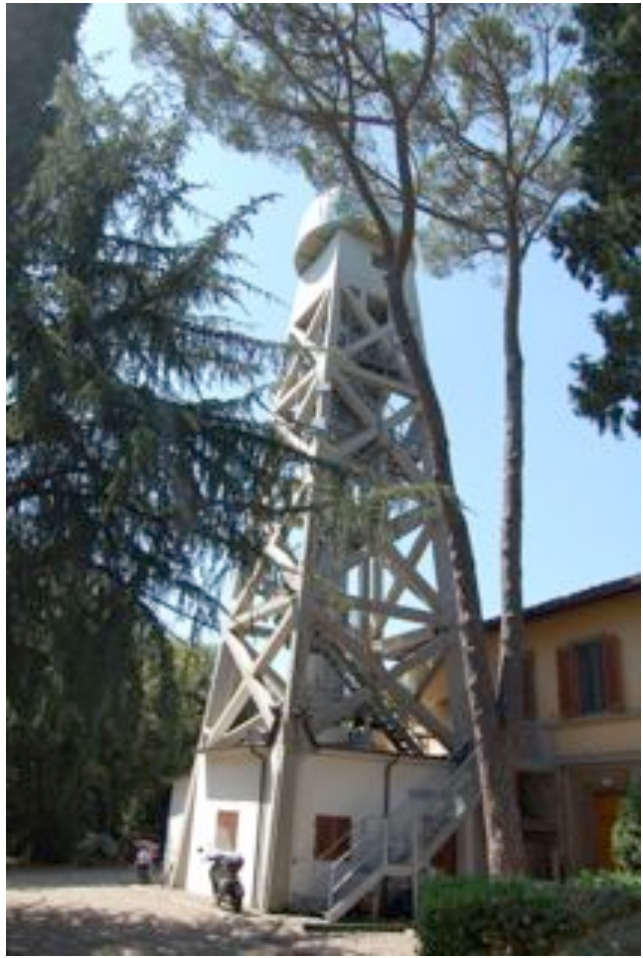

**Figure 15:** The ASM test setup in the Arcetri test tower. The concave secondary allows for a relatively simple double-pass optic test configuration at the same focal length that it will be used at the telescope. A calibration source in the NAS at the far focus reflects off of the ASM, then the CRO, then back to the CRO before entering the W-unit for wavefront sensing and technical viewing.

### 3.2 CRO Fiducials and Alignment

Like the ASM and PWFS, much of the test setup at the Arcetri test tower was identical to that used to test the LBT AO systems. However, one major difference is that the LBT ASM is mounted to a hexapod that can be used to align the ASM to a stationary CRO. In the test tower setup, the ASM surface must be aligned to both of its ellipsoidal conjugates, i.e. the CRO and the fiber point source. The Magellan telescope does not provide a hexapod to move the ASM w.r.t. the CRO. Therefore, we designed a new frame and mount to allow us to move the CRO to a stationary ASM.

Because of the ~16:1 focal ratio of the near and far ellipsoidal conjugates of the ASM, the CRO alignment is extremely sensitive and must be aligned to within a micron or two of the ASM conjugate. To that end, we mounted the CRO to a precision 5-axis piezo actuator stage that allows to reach this precise alignment. We also wanted to build in the capability of being able to go on and off sky quickly if needed during commissioning. To achieve this, we designed a repeatable ball-and-groove magnetic kinematic interface between the CRO mount plate and the mounting truss. When the CRO is removed, light is allowed to pass through focus at the near ASM conjugate. Therefore, once the CRO alignment is achieved once, we can remove it and replace it easily when we want to go back and forth between a real star and the off-sky fiber source test.

One further complication to our CRO scheme was that the 5-axis piezo stage has a very limited travel range of only a few mm. If the CRO is not already aligned to within a mm or two when the CRO is installed, then the alignment cannot be achieved. During the shipping from Arcetri to Chile, the entire truss assembly was removed from the ASM wind screen, with several bolted interfaces removed that were not designed to necessarily be repeatable upon reassembly. Therefore, in order to maintain some fiducial of the precision CRO alignment that we had achieved in the Arcetri test tower, we designed a very compact mount for a small (U.S. quarter-sized) handgun laser. This laser repeatably mounts to a delrin mounts surface at the very center of the ASM itself, in the region shadowed by the telescope's central obscuration. The laser beam fires straight down the optical axis from the center of ASM, where the vertex would be, to the short ellipsoidal conjugate. When the CRO is in place, the laser enters to the entrance to the CRO cup. At the telescope during commissioning, the laser will also be used as a convenient boresight reference during the installation of the ASM.

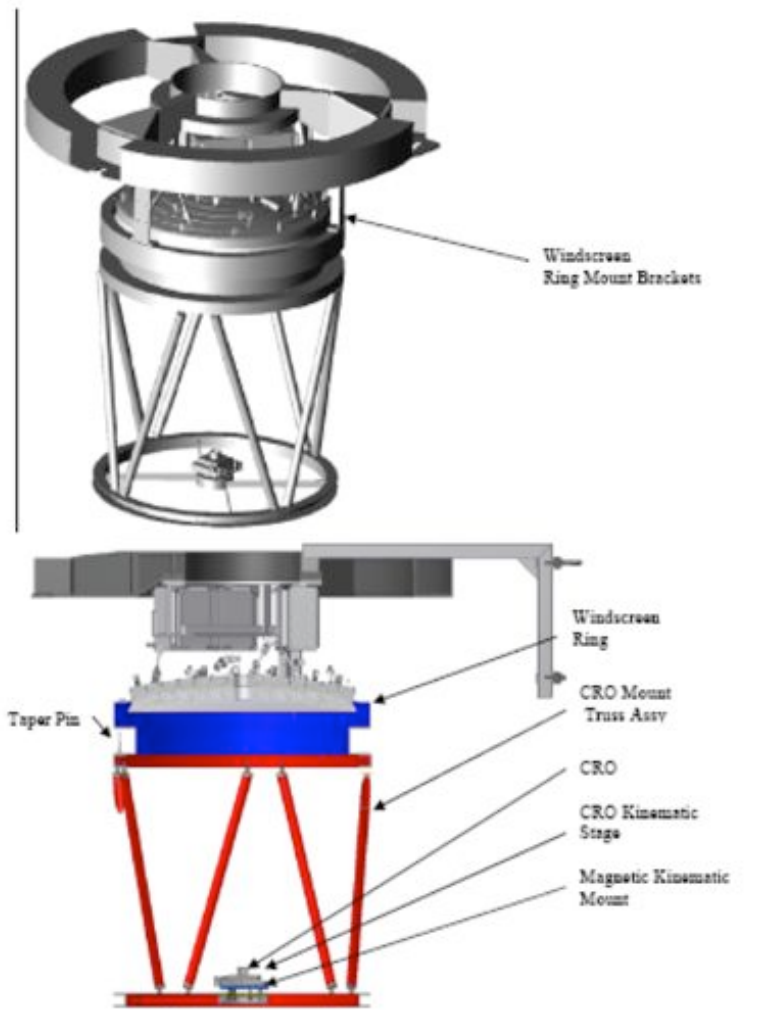

**Figure 16:** The ASM and CRO truss assembly. This setup is unique to the Magellan system and contains a remotely actuated CRO that can be aligned to the ASM optical focus to micron-level precision.

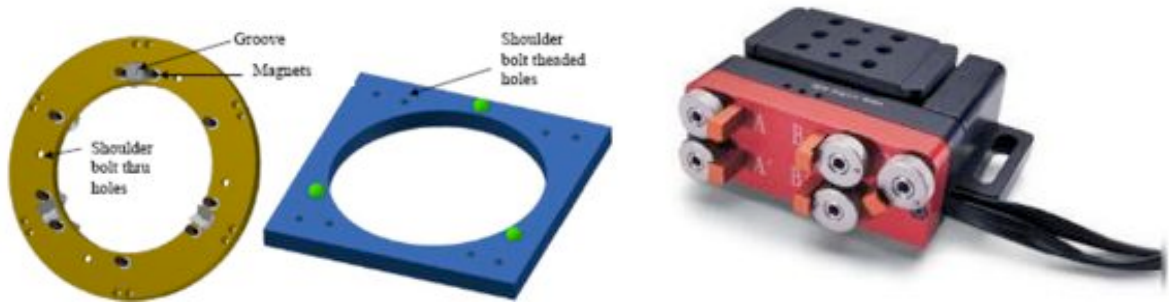

**Figure 17:** Left: The magnetic kinematic interface plates that allow the CRO to be easily and repeatably removed when switching between the off-sky and on-sky configurations. Right: The 5-axis piezo stage for aligning the CRO to the ASM.

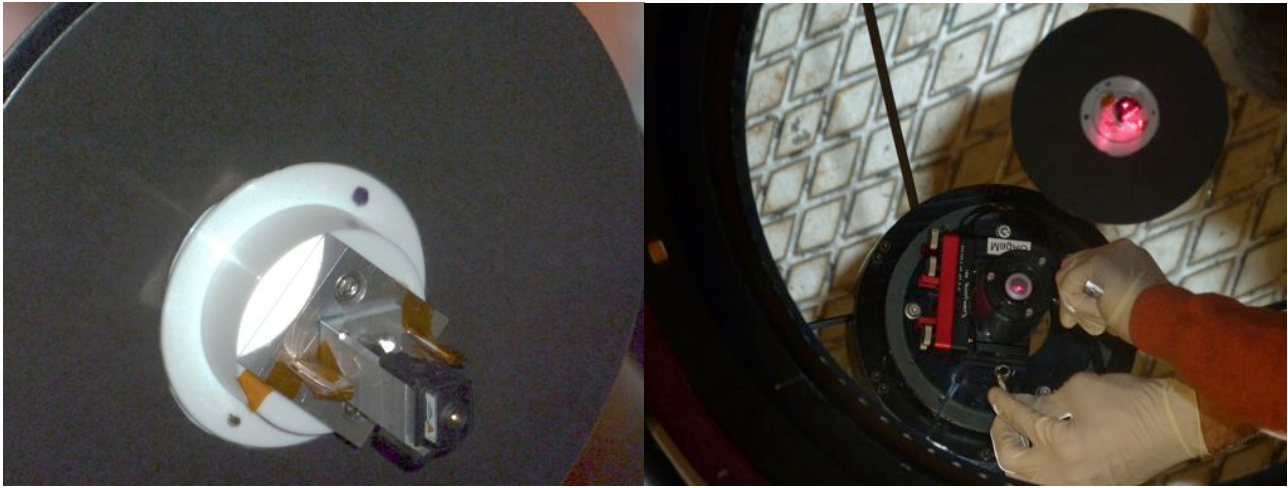

**Figure 18:** Left: The rail mounted handgun laser mounted to a delrin cap at the center of the ASM. The laser acts as a optical axis boresight reference and is also used to verify the location of the CRO. Right: The CRO cup mounted to its piezo stage and magnetic kinematic mount. The boresight laser is shining directly into the hole in the CRO return flat, indicating a good alignment. This view is the reflection in the ASM.

## 4. RESULTS

We tested the full AO system in the test tower in a wide variety of wavelengths, levels of turbulence, PWFS binning, etc. For more on the AO tower testing, see Males et al. these proceedings. The full AO system with all of the optical components, including the ADC in its non-dispersive orientation, performs as predicted with excellent image quality at $\lambda$ = 982 nm and Strehls of 85% (see Males et al. 2012 for more discussion on the turbulence simulation in this result). This is excellent performance and should allow us to meet our VisAO science goals.

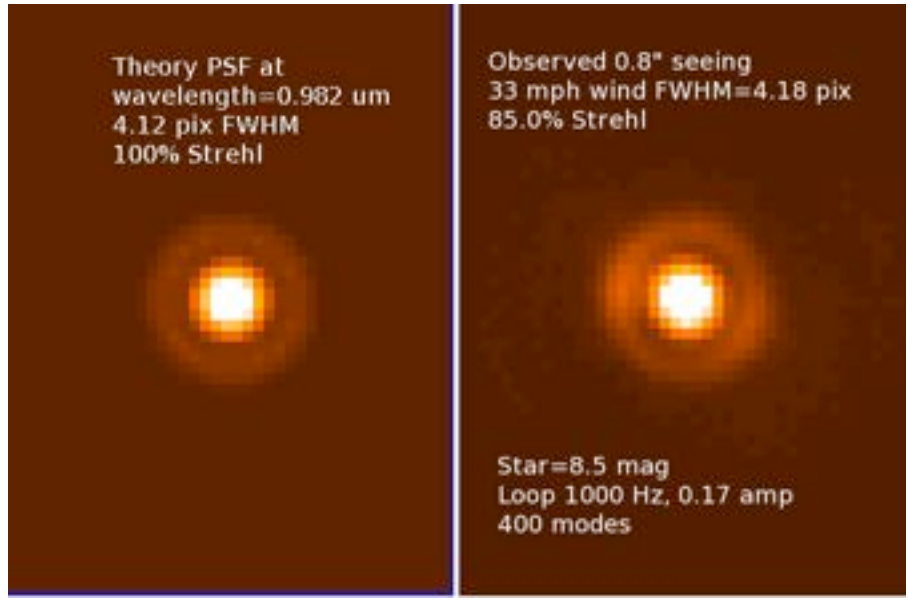

**Figure 19:** Closed loop performance at 982 nm of the complete AO system in the Arcetri test tower. Left: Theoretical PSF at this F/# and wavelength. Right: AO corrected image of the fiber source with the tower vibration, ambient environment, and injected "seeing" as the sources of turbulence.

## 5. CONCLUSION

The Magellan AO system, recently having passed a successful pre-ship review in Florence, Italy, is scheduled for first light on the Magellan Clay telescope in November 2012. The excellent seeing of the Magellan site, combined with a high-performance adaptive secondary AO system whose capabilities have already been demonstrated with the virtually identical LBT system and with our own tower tests, will provide unprecedented high-resolution images of targets in the southern sky in both the visible and mid-IR bands. The VisAO camera will demonstrate many novel techniques for achieving this high-resolution in the visible, many of which we discussed in this paper. These include the advanced triplet ADC, the fast asynchronous shutter for real-time frame selection, the calcite Wollaston prism and corresponding filters for SDI imaging, and our custom chrome-on-glass coronographic spots.

## ACKNOWLEDGEMENTS

This project owes a debt of gratitude to our partners and collaborators. The ASM and WFS could not have been possible without the design work of Microgate and ADS in Italy as well as Arcetri Observatory and the LBT observatory. We would like to thank the NSF MRI and TSIP programs for generous support of this project in addition to the Magellan observatory staff and the Carnegie Institute. We would also like to thank the engineers and staff at Optimax for doing an excellent job fabricating the ADC, a challenging and unique optic.